\documentstyle[aps,preprint]{revtex}
\begin{document}
\draft
\title{Flux Noise near the
  Berezinskii-Kosterlitz-Thouless Transition}
\author{Karl-Heinz Wagenblast$^{a)}$ and Rosario Fazio$^{b),c)}$}
\address{$a)$Institut f\"ur Theoretische Festk\"orperphysik,
  Universit\"at Karlsruhe, D-76128 Karlsruhe, Germany}
\address{$b)$Istituto di Fisica, Universit\`a di Catania,
  viale A. Doria 6, I-95129 Catania, Italy}
\address{$b)$Istituto Nazionale di Fisica della Materia (INFM), Unit\`a di
  Catania, Italy}

\maketitle

\begin{abstract}
We study the flux noise in Josephson junction arrays in the critical
regime above the Berezinskii-Kosterlitz-Thouless transition.  In proximity
coupled arrays a local ohmic damping for the phases is relevant, giving
rise to anomalous vortex diffusion and a dynamic scaling of the flux
noise in the critical region.  It shows a crossover from white to
$1/f$-noise at a frequency $\omega_\xi\propto\xi^{-z}$ with a dynamic
exponent $z=2$.  
\end{abstract}

A variety of two dimensional systems undergo the
Berezinskii-Kosterlitz-Thouless (BKT) pair-unbinding
transition~\cite{berezinskii,kosterlitz}. The transition
between the high temperature (disordered) and the low temperature
(coherent) phase is driven by the thermally excited vortices. These
topological excitations form a two dimensional Coulomb gas. Below the
BKT transition temperature $T_{\text{BKT}}$ they are bound in
vortex-antivortex pairs. Above the critical temperature the pairs
dissociate and form a vortex plasma. In the plasma phase the
vortex-vortex interaction is screened at a distance of the order of
the correlation length $\xi$ which diverges at $T_{\text{BKT}}$.

Arrays of Josephson junctions are prototype systems to study the BKT
transition. Below the BKT transition the array is phase
coherent and thus superconducting.  On the contrary, in the plasma
phase, free vortices destroy the coherence and the system is
resistive, though each island remains superconducting.  In the last
decade there has been a great amount of work on the various aspects of
the BKT transition in Josephson arrays (see Ref.~\cite{Mooij&Schoen} for
an overview). Experimental studies are based on electrical
resistance~\cite{exp-resistance}, two-coil
inductance~\cite{exp-inductance,theron}, and SQUID~\cite{shaw,lerch}
measurements.

Anomalous suppression of the vortex mobility has been observed in
proximity coupled Josephson junction arrays by Th\'eron
{\em et al.}~\cite{theron}. The vortex mobility vanishes logarithmically
at low energies, showing that vortex motion in these arrays
{\em  cannot} be described by a Drude model with a constant mobility.
This behavior has been explained either introducing a local
ohmic damping for the phases of the superconducting wave
function~\cite{beck,korshunov} or by invoking long-range
vortex interactions~\cite{houlrik}. More recently a regime interpretation
of the vortex dynamics was given in Ref.\cite{Capezzali97}

In a recent Letter, Shaw {\em et al.}~\cite{shaw} investigated the
magnetic flux noise close to the BKT transition in an overdamped
Josephson junction array by means of a SQUID.
The flux-noise spectrum is defined as
\begin{equation}
  S_\Phi(\omega) = \int dt e^{i\omega t}\langle \Phi(t) \Phi(0)\rangle\;,
\end{equation}
where $\Phi(t)$ is the flux detected by the SQUID and is proportional
to the vortex density integrated over the SQUID area. Near the
transition the flux noise is white for frequencies $\omega<\omega_\xi$
crossing over to a $1/f$-noise for $\omega>\omega_\xi$. The crossover
frequency $\omega_\xi$ vanishes as the transition is approached from
above as $\omega_\xi \propto \xi^{-z}$, where $\xi$ is the correlation
length and $z\approx 2$ is the dynamic exponent. All the data obtained
at different temperatures collapse on a single scaling curve when
plotted as a function of $\omega/\omega_\xi$. The problem was studied by
numerical simulations~\cite{Tiesinga97,Hwang97}  of both the Time Dependent
Ginzburg-Landau (TDGL) and the Resistive Shunted Junction (RSJ) models.

In this Letter we study theoretically the flux noise in the TDGL 
model near the BKT transition. We determine the scaling function of the 
flux noise analytically in the limit in which the SQID size is much smaller 
than the correlation lenght.

In order to study the flux noise we map the XY-model with a local
damping onto a 2D (dynamic) Coulomb gas~\cite{minnhagen}. We determine 
the dynamic correlation functions using an improved Debye-H\"uckel 
approach for the Coulomb gas in the same spirit as in Ref.~\cite{berker}.

The Euclidean action for the XY-model with local damping for the
phases $\varphi$ is
\begin{equation}
  S[\varphi] = E_J \int_0^{\beta} d\tau \sum_{<ij>}
  \left[ 1-\cos(\varphi_i(\tau)-\varphi_j(\tau))\right]
   +\frac{\alpha}{8\beta^2}\int_0^{\beta} d\tau d\tau' \sum_i 
  \left(\frac{\varphi_i(\tau)-\varphi_i(\tau')}
    {\sin[\pi(\tau-\tau')/\beta]}\right)^2\,,
\end{equation}
where $E_J$ is the Josephson coupling between neighboring sites. The
local ohmic damping introduces an interaction of the phase of a single
grain at different times. In proximity coupled arrays, which consist
of superconducting island on top of a metallic film, the local damping
parameter $\alpha$ is related to the shunting resistance $R$ of one
island to the substrate, $\alpha=h/4e^2R$.

A Villain transformation~\cite{villain} is used to derive an effective
dynamic action for the vortices. Here we follow the method described in
Ref.~\cite{fazio} and obtain the dynamic action for the vortex degrees
of freedom $v_i(\tau)$
\begin{equation}
  S[v]=\frac{1}{2}\int_0^{\beta} d\tau d\tau' \sum_{i,j}
  v_i(\tau) D_{ij}(\tau-\tau') v_j(\tau')\,.
\end{equation}
The kernel $D$ reads
\begin{equation}
  D(k,\omega_{\mu})=\frac{4\pi^2 E_J}{k^2}
  \frac{k^2+2 |\omega_\mu|/\omega_\alpha}
  {k^2+|\omega_\mu|/\omega_\alpha}\,,
\end{equation}
with $\omega_\alpha=2\pi E_J/\alpha$, and $\omega_{\mu}$ are the
Matsubara frequencies.  We choose the lattice spacing as the unit of
length scales. In the static case this yields the standard 2D Coulomb gas.

The evaluation of the flux noise is done by means of a two-step
procedure in the spirit of Ref.~\cite{berker}.  
i)The physics taking place at length scales shorter than $\xi$ 
is taken into account by using the scaling properties of the flux noise.
ii) At the scale of the order of the correlation lenght, all the dipoles
have been integrated out and, since the physics above $T_{\text{BKT}}$ is 
dominated by screening, a Debye-H\"uckel approximation (with renormalized
parameters) is used~\cite{halperin,berker}.  

The effect of bound pairs up to distances of the order of the correlation
length can be account for by means of the scaling behaviour of the flux noise
\begin{equation}
  \label{scaling}
  S_{\Phi}(\omega,l,T')=\text{e}^{z\delta}
  S_{\Phi}(\omega\text{e}^{z\delta},l\text{e}^{-\delta},T'(\delta))\,.
\end{equation}
At $\delta^*=\ln (\xi/\xi_0)$ all the vortices are integrated out
up to a distance of the order of $\xi$.  At this scale only free
vortices are present and the Debye-H\"uckel approximation can be used
to calculate the r.h.s.\ of Eq.~(\ref{scaling}).

The effect of screening due to the presence of free
vortices above $T_{\text{BKT}}$ is analyzed in the Debye-H\"uckel approximation.
The Matsubara Green's function for the vortices in this approximation is
\begin{equation}
  G(k,\omega_\mu)=\langle vv\rangle_{k,\omega_\mu}
  =\frac{1}{4\pi^2E_J\xi^2+D(k,\omega_\mu)}
\label{Debye-Huckel}
\end{equation}
The Kosterlitz correlation length $\xi$ diverges exponentially near the transition,
$\xi=\xi_0\exp(b/\sqrt{T'-T'_{\text{BKT}}})$, with $T'=T/E_J$
\cite{kosterlitz,shaw,olsson}. An analytic continuation ($|\omega_\mu|
\rightarrow -i\omega$) yields the retarded vortex propagator
$G^R(k,\omega)=G(k,|\omega_\mu| \rightarrow -i\omega)$ which is
related to the spectral function of vortex density-density
correlations by the fluctuation-dissipation
theorem $\langle vv\rangle_{k,\omega }= {\mbox{Im}}\,G^R
(k,\omega)\, 2T/\omega$, for $\omega \ll T$. This correlation
function describes the equilibrium vortex fluctuations in the system
and gives rise to the flux noise.

Experiments on the magnetic flux noise detect the fluxes of the
vortices from an effective area $l^2$~\cite{theron,shaw,lerch}.
The relevant quantity is the vortex density fluctuation
integrated over the pick-up area $\; \Phi (t) = \int_ld^2x \; v(\vec{x},t)$. 
After having performed the analytic continuation and by combining 
the scaling arguments (see Eq.~(\ref{scaling})) with the Debye-H\"uckel 
approximation (see Eq.~(\ref{Debye-Huckel})), we obtain the flux-noise spectrum
\begin{equation}
  \label{sphi}
  S_\Phi(\omega) = \Phi_0^2\ \text{e}^{z\delta^*}S_{\Phi,\text{DH}}
  (\omega\text{e}^{z\delta^*},l\text{e}^{-\delta^*},T'(\delta^*))
  =\frac{{\cal C}\;\Phi_0^2}{8\pi^3}\frac{T'(\delta^*)l^4}{\omega \xi^4} 
  F\left(x=\frac{\omega}{\omega_\xi},y=\frac{4\pi\xi^2}{l^2}\right)\,,
\end{equation}
where the subscript DH means at at this scale the flux noise can be evaluated by means
of the Debye-H\"uckel approximation.
The constant $\cal C$ takes into account the geometrical 
details of the experimental setup. We introduced the scaling function $F$
\begin{equation}
  \label{sca1}
  F\left(x=\frac{\omega}{\omega_\xi},y=\frac{4\pi\xi^2}{l^2}\right)
  =-{\mbox{Im}}{\int_0^y dz
  \left[1+z\frac{z-ix}{z-2ix}\right]^{-1}}\!,
\end{equation}
where characteristic frequency is $\omega_\xi=\omega_0 (\xi_0/\xi)^2$
(we used a hard cutoff in the $k$-space to integrate over the pick-up
coil area). The dynamic exponent $z=2$ directly emerges. The
frequency scale $\omega_0$ is related to microscopic parameters,
$\omega_0=2\pi E_J/(\xi_0^2\alpha)$. The scale of the correlation
length, $\xi_0$, is of the order of the lattice spacing.
Close to the transition ($\xi\rightarrow\infty$) the scaling function
$F$ reduces to
\begin{equation}
  \label{sca2}
  F(x,y\rightarrow\infty) =\mbox{Im}\frac{1+3ix}{2Q(x)}
  \ln\left(\frac{1-ix+Q(x)}{1-ix-Q(x)}\right)
  -\frac{\pi}{4}
\end{equation}
where $Q(x)^2=1-x^2+6ix$. 
The flux noise in the relevant limits is given by
\begin{equation}
  \omega S_{\Phi} (\omega) \sim \left\{ \begin{array}{ll}
      	\frac{{\cal C}\;\Phi_0^2}{8\pi^3}\frac{T'(\delta^*)l^4}{\xi^4}     
	& \mbox{for $\frac{\omega}{\omega_\xi} \gg 1$}\\
      	\frac{{\cal C}\;\Phi_0^2}{8\pi^3}
	\frac{T'(\delta^*)l^4}{\xi^4} \frac{\omega}{\omega_\xi}
	& \mbox{for $\frac{\omega}{\omega_\xi} \ll 1$}\,,
    \end{array}
  \right.
\end{equation}
The white noise occurs for frequencies
$\omega<\omega_\xi$, and $1/f$-noise for $\omega>\omega_\xi$. The $1/f$-noise
stems from the superposition of Lorentzian-shaped contributions at all 
length scales down to the dimension of the SQUID.  For finite $\xi$ 
the $1/f$ noise crosses over to a $1/f^2$-noise. The $1/f$-noise is
intermediate in the frequency range $1<\omega/\omega_\xi<4\pi\xi^2/l^2$.  
Close to the transition this range exceeds several orders of magnitude.

We finally apply our results to the experiment
by Shaw {\em et al.}~\cite{shaw}. The critical coupling in the
experiment is $T'_{\text{BKT}}=0.06$, thus the bare fugacity at
criticality is $y_0=0.35$. Assuming that the bare fugacity is only
weakly temperature dependent, the numerical iteration of the RGE gives
to a good approximation $T'(\delta)\propto\exp(4\delta)$, as upon
renormalization the limit $T'(\delta) \gg 2/\pi$ is reached, where
$y(\delta)\propto\exp(2\delta)$ (this scaling behavior reflects the
fact that the vortex-pair density is constant). Thus
$T'(\delta^*)/\xi^4$ is temperature independent and we can identify
our scaling function $F$ with the experimental scaling behavior.
These last considerations should be considered more as additional
assumptions whose validity is mostly based in the comparison with the
experiments which can be made quantitative.  
The characteristic frequency was found to be $f_0=2.1\times 10^6\,$Hz. 
With the estimate $E_J=10$K we can extract $\xi_0^2\alpha=10^6$,
corresponding to ohmic shunts to the ground with a resistance 
$R=6\,$m$\Omega\times\xi_0^2$. 
We expect $\xi_0$ to be of the order of one lattice spacing. Yet another
possibility is to determine the resistance $R$ and deduce the value of
$\xi_0$. The effective shunting resistance of one island to infinity
can be estimated be of the order of $10\,$m$\Omega$.  This also
yields the scale of the correlation length to be of the order of one
lattice spacing.

In summary we calculated the flux noise above the BKT transition and
derived its scaling behavior analytically in the case where the
correlation length exceeds the size of the pick-up loop. We find a
crossover from white to $1/f$-noise.  We related the crossover
frequency to microscopic parameters of the array.  Our results may explain
a recent experiments by Shaw {\em et al.}~\cite{shaw}.

We are indebted to T.J.~Shaw for useful conversations and for helping
us in comparing our work with the experiment. We would like to thank
J.~Clarke, D.-H.~Lee, P.~Minnhagen, Gerd Sch\"on, A.~Schmid, and
M.~Tinkham for fruitful discussions. This work was supported by
``SFB 195'' of the ``Deutsche
Forschungsgemeinschaft'' and the EU-grant TMR-FMRX-CT-97-0143.

\end{document}